\documentclass[aps,pra,twocolumn,american,showkeys,showpacs,floatfix,reprint]{revtex4-1}
\usepackage{lmodern}
\usepackage[T1]{fontenc}
\usepackage[koi8-r,latin9]{inputenc}
\usepackage{geometry}
\usepackage{units}
\usepackage{amsmath}
\usepackage{amssymb}
\usepackage{graphicx}
\usepackage{setspace}

\makeatletter
 
 \@ifundefined{textcolor}{}
 {%
   \definecolor{BLACK}{gray}{0}
   \definecolor{WHITE}{gray}{1}
   \definecolor{RED}{rgb}{1,0,0}
   \definecolor{GREEN}{rgb}{0,1,0}
   \definecolor{BLUE}{rgb}{0,0,1}
   \definecolor{CYAN}{cmyk}{1,0,0,0}
   \definecolor{MAGENTA}{cmyk}{0,1,0,0}
   \definecolor{YELLOW}{cmyk}{0,0,1,0}
 }

\makeatother

\usepackage{babel}
\begin{document}

\title{COALESCENT PHOTONS IN BLACKBODY RADIATION}

\author{Aleksey Ilyin}

\email{a.v.ilyin@mail.mipt.ru}

\affiliation{Moscow Institute of Physics and Technology}

\date{\today}
\begin{abstract}
Within the Bose-Einstein statistics it is shown that the blackbody
radiation contains coalescent photons along with single photons. Coalescent
photons were first observed in the famous Hong\nobreakdash-Ou\nobreakdash-Mandel
experiment of 1987. For the sake of convenience, $N$ coalescent photons
are referred to as the $N$\nobreakdash-photon cluster. In this work,
statistics of photon clusters and probability that a photon cluster
contains $N$ photons are found versus radiation frequency and temperature.
Spectra of photon\nobreakdash-cluster radiation are calculated as
functions of blackbody temperature for different cluster ranks. Derivation
of the Planck's radiation law is discussed in view of the existence
of photon clusters in blackbody radiation.
\end{abstract}

\keywords{Bose-Einstein statistics, thermal radiation, Planck's formula, coalescent
photons}

\pacs{14.70.Bh; 42.50.-p; 44.40.+a}

\maketitle

\section{INTRODUCTION}

In this work we consider yet unknown properties of blackbody radiation,
which are derived here from the Bose-Einstein (BE) statistics. It
is proven that a part of blackbody radiation energy is carried by
coalescent photons.

Coalescent photons were first discovered by Hong, Ou and Mandel in
their seminal experiment~\cite{HOM}. Since then, physics of coalescent
photons has attracted growing attention fostered by the possibility
of using coalescent photons in quantum information science and in
the tests of fundamental concepts of quantum physics~\cite{Bouwmeester,M_Genovese,Quantum Computing}.

The properties of coalescent photons are usually analyzed theoretically
on the basis of wave functions~\cite{Ou_Mandel_Review,Yanhua Shih-2,Kaige Wang}.
A~new approach to this problem proposed in this work is based on
quantum statistics and provides new information on the physics of
coalescent photons in blackbody radiation. 

In this paper, $K$ coalescent photons are referred to as a $K$\nobreakdash-photon
cluster or a photon cluster of rank~$K$. Such terminology is justified
because two coalescent photons were found to interact with a beamsplitter
as if they were a single quantum object~\cite{Stability_of_clusters}.
In other words, this single quantum object, or a two-photon cluster,
is either transmitted through the beam splitter, or reflected from
it as a whole. 

It is shown in Sections~\ref{sec:Basic ideas}\nobreakdash-\ref{sec:Cluster radiation spectra}
that the BE statistics predicts the existence of photon clusters in
thermal radiation and determines both the cluster statistics and radiation
spectra for clusters of various ranks in a blackbody cavity. 

\begin{figure}[ht]
\noindent \begin{centering}
\includegraphics[width=8.5cm]{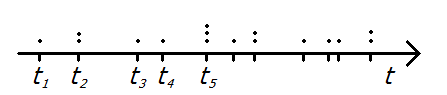}
\par\end{centering}

\caption{\label{fig:CPD}\small An example of discrete random process described by
the \emph{Compound Poisson Distribution}. Discrete events occur at
random moments of time $t_{1}$, $t_{2}$, $\ldots$ $t_{n}$, $\ldots$
Some events shown in the figure, like those of $t_{1}$, $t_{3}$
and $t_{4}$, are elementary (or single) events, while there are also
compound events, like the double{\small \protect\nobreakdash-}event
at $t_{2}$, or triple{\small \protect\nobreakdash-}event at $t_{5}$
that consists of three elementary events occurring at the same time.}
\end{figure}

The existence of photon clusters in thermal radiation is a formal
consequence of the fact that the BE statistics is a negative binomial
distribution (NBD), which is known to be a special case of \emph{Compound
Poisson Distribution}~\cite{Compound_Distribution}. That means that
the BE statistics describes random events of two different types:
elementary events and compound events, each compound event consisting
of random number of elementary events. An example of random process
described by a Compound Poisson Distribution is shown in Fig.~\ref{fig:CPD}.

In the case of photon statistics, an elementary event may only be
the registration of one photon by an ideal detector. Then a compound
event will be a simultaneous registration of several photons, i.\,e.
a photon cluster. 

Basic ideas required to understand further results are introduced
in Section~\ref{sec:Basic ideas}, which explains why the BE statistics
predicts the existence of photon clusters. 

In Section \ref{sec:FS} and Appendix~1, a formal proof is given
that photon clusters are inherent to the BE statistics. Section \ref{sec:Composition of Distributions}
shows that the BE statistics is the result of superposition of two
random processes, first, a random number of photon clusters entering
a given phase-space volume; second, a random number of photons contained
in each cluster. Such superposition of two random processes is known
as a superposition of distributions.

Radiation spectra of photon clusters of various ranks in the blackbody
cavity are found in Section \ref{sec:Cluster radiation spectra}.
It is shown that the sum of cluster radiation spectra taken over all
cluster ranks (from unity to infinity) yields the Planck's radiation
law for blackbody radiation spectrum as a function of frequency ($\nu${\small \nobreakdash-}spectrum).
Thus, it is shown that clusterization of photons does not affect the
$\nu${\small \nobreakdash-}spectrum of blackbody radiation.

In contrast to the $\nu${\small \nobreakdash-}spectrum, the $\lambda${\small \nobreakdash-}spectrum
of blackbody radiation is found to be sensitive to the process of
photon clusterization, which shall inevitably result in some modification
of the Planck's radiation law for the $\lambda${\small \nobreakdash-}spectrum
of blackbody radiation. However, the actual degree of clusterization,
as shown in Section \ref{sec:Cl_spectra_vs_lambda}, leads to a minor
modification of the Planck's radiation law not exceeding 7.5\% in
its maximum for reasonable temperatures of blackbody radiation. 

Main results of this work are discussed in Conclusions (Section~\ref{sec:Conclusions}).

\section{BASIC IDEAS\label{sec:Basic ideas}}

In this Section we consider classical and quantum probabilities \foreignlanguage{english}{$p_{n}(\tau)$}
that $n$ particles are in some volume~$\tau$. All phase-space volumes
are assumed to be measured in the units of $\hbar^{3}$ while three-dimensional
volumes are measured in coherence volumes. Therefore, all volumes
in this article are dimensionless.

\subsection{Simplified classical approach}

In the classical case, the probability that $n$ non-interacting particles
are in volume $\tau$ is determined by the Poisson statistics: 
\begin{equation}
p_{n}(\tau)=\dfrac{(w\tau)^{n}}{n\,!}\, e^{-w\tau},\label{eq:Classic}
\end{equation}
where $\omega$ - is the average number of particles per unit volume.
According to \eqref{eq:Classic}, the probability that no particles
are in volume $\tau$ is
\begin{equation}
p_{0}(\tau)=e^{-\omega\tau},\label{eq:P0_cl}
\end{equation}
therefore, \eqref{eq:Classic} can be written as 
\begin{equation}
p_{n}(\tau)=\dfrac{(w\tau)^{n}}{n\,!}\, p_{0}(\tau).\label{eq:Classic_using_P0}
\end{equation}

In some theoretical considerations it is convenient to deal with relative
probabilities defined as
\begin{equation}
q_{n}(\tau)=\dfrac{p_{n}(\tau)}{p_{0}(\tau)},\label{eq:Rel_Prob_q}
\end{equation}
which for Poisson statistics \eqref{eq:Classic_using_P0} yields 
\begin{equation}
q_{n}(\tau)=\dfrac{(w\tau)^{n}}{n\,!}.\label{eq:Rel_Prob_in_Poisson}
\end{equation}
In this case $q_{n}(\tau)$ will be proportional to~$\tau^{n}$ for
any volume~$\tau$. For example, relative probability that one particle
is in an arbitrary volume~$\tau$
\begin{equation}
q_{1}(\tau)\sim\tau,\label{eq:1P}
\end{equation}
while the probability that two particles are in the same volume 
\begin{equation}
q_{2}(\tau)\sim\tau^{2},\label{eq:2P}
\end{equation}
which is natural because classical particles enter the volume independently
of each other. 

Now let us consider a gas of classical particles that may stick together
with some nonzero probability. In this case, two particles stuck together
will form a new object that may be termed a two-particle ``molecule''
or a two-particle ``cluster''. Such molecules, like separate particles,
will enter the selected volume independently of each other. Therefore,
the probability that $n$ molecules are in volume $\tau$ must obey
Poisson statistics \eqref{eq:Classic} but with a different value
of molecule density~$w$. 

Consequently, relative probability that one molecule is in volume
$\tau$ should be proportional to $\tau$ by analogy with~\eqref{eq:1P}.
For this reason, if two \emph{coalescent} particles are in the selected
volume then relative probability of such event is proportional to
$\tau$, while the same probability will be proportional to~$\tau^{2}$
if the two particles are separate. Therefore, relative probability
$q_{2}(\tau)$ that two particles are in volume $\tau$ in the general
case should be 
\begin{equation}
q_{2}(\tau)\thicksim k_{1}\tau+k_{2}\tau^{2},\label{eq:Two_particles}
\end{equation}
where the first term describes the probability that a two-particle
molecule is in volume $\tau$ while the second term is responsible
for the probability that two separate particles are in the said volume.

If two particles are stuck together then they are not independent.
Therefore, formula \eqref{eq:Two_particles} signifies a departure
from the classical Poisson statistics \eqref{eq:Classic}, which is
applicable to independent particles only. The polynomial character
of relative probability \eqref{eq:Two_particles} as a function of
volume indicates that some particles are stuck together and exist
in the form of clusters, or molecules, while other particles remain
separate.

\subsection{Simplified quantum approach}

It is easy to show that quantum statistics yields for relative probability
$q_{2}(\tau)$ a non-classical dependence against volume similar to
eq.~\eqref{eq:Two_particles}.

Indeed, the BE statistics is usually considered for a single cell
of phase space
\begin{equation}
p_{n}(1)=\dfrac{w^{n}}{\left(1+w\right)^{n+1}},\label{eq:BE_in_single_bin}
\end{equation}
where $p_{n}(1)$ is the probability that $n$ photons are in one
coherence volume, $w$ is the average number of photons per coherence
volume. Designating quantum and classical probabilities by the same
symbol $p_{n}$ may cause no confusion. 

For arbitrary volume $\tau$, the BE statistics has the form:
\begin{equation}
p_{n}(\tau)=C_{\tau+n-1}^{n}\frac{w^{n}}{\left(1+w\right)^{n+\tau}},\label{eq:BE_in_tau_bins}
\end{equation}
where 

\begin{eqnarray}
C_{\tau+n-1}^{n}=\dfrac{\left(\tau+n-1\right)\,!}{n\,!\left(\tau-1\right)\,!} & = & \frac{\tau(\tau+1)\ldots(\tau+n-1)}{n\,!}.\nonumber \\
\label{eq:Comb}
\end{eqnarray}

Formula \eqref{eq:BE_in_tau_bins} was derived by Leonard Mandel for
an integer number of cells~\cite{BE_by_Mandel}. It was later shown
in \cite{GBD} that the Mandel's formula \eqref{eq:BE_in_tau_bins}
is valid for an arbitrary volume~$\tau$ including nonintegral number
of coherence volumes. It is clear that the last expression in \eqref{eq:Comb}
makes sense for any positive volume~$\tau>0$. If $\tau=1$ then
\eqref{eq:BE_in_tau_bins} becomes the usual expression \eqref{eq:BE_in_single_bin}
for the BE statistics in a single cell.

From \eqref{eq:BE_in_tau_bins} we obtain the probability that no
particles are in volume~$\tau$ 
\begin{equation}
p_{0}(\tau)=\dfrac{1}{\left(1+w\right)^{\tau}}.\label{eq:P0_BE}
\end{equation}
Here we have taken into account that if $n=0$ then $C_{\tau+n-1}^{n}=1$
for any~$\tau$.

Given \eqref{eq:BE_in_tau_bins} and \eqref{eq:P0_BE}, relative probability
that $n$ quantum particles are in volume~$\tau$, by analogy with~\eqref{eq:Rel_Prob_q},
can be written as 
\begin{equation}
q_{n}(\tau)=\dfrac{p_{n}(\tau)}{p_{0}(\tau)}=\dfrac{\tau(\tau+1)\ldots(\tau+n-1)w^{n}}{n\,!\left(1+w\right)^{n}}.\label{eq:Rel_Prob_in_BE}
\end{equation}
Hence, the relative probability that two particles are in volume~$\tau$
is
\begin{equation}
q_{2}(\tau)=\dfrac{\tau(\tau+1)w^{2}}{2\,!\left(1+w\right)^{2}}\thicksim c_{1}\tau+c_{2}\tau^{2},\label{eq:Two_ph_in_BE}
\end{equation}
where $c_{1}$ and $c_{2}$ are some coefficients. Eq.~\eqref{eq:Two_ph_in_BE}
for quantum particles is of the same form as eq.~\eqref{eq:Two_particles}.
Therefore, the volume dependence of relative probability \eqref{eq:Two_ph_in_BE}
in the BE statistics suggests that some particles may stick together
while other particles remain separate.

The simplified analysis presented above is based on an intuitive understanding
that if some particles are stuck together then this fact must influence
relative probability \eqref{eq:Two_ph_in_BE} as a function of volume.
In the following Section, a formal solution to this problem is discussed
based on the properties of Compound Poisson Distribution as presented
in~\cite{Compound_Distribution}.

\section{FORMAL SOLUTION IN QUANTUM STATISTICS\label{sec:FS}}

Coalescence of particles in BE statistics is a consequence of the
fact that the BE statistics coincides with a negative binomial distribution,
which has the form
\begin{equation}
p_{n}(\tau)=C_{\tau+n-1}^{n}p^{\tau}\left(1-p\right)^{n},\label{eq:NBD}
\end{equation}
where coefficients $C_{\tau+n-1}^{n}$ are defined in
\eqref{eq:Comb}, parameters $p$ and $\tau$ must satisfy $0\leq p\leq1$
and $\tau>0$, respectively. If parameter 
\begin{equation}
p=\dfrac{1}{1+w}\label{eq:p(w)}
\end{equation}
then \eqref{eq:NBD} coincides with the BE statistics~\eqref{eq:BE_in_tau_bins}.

According to \cite{Compound_Distribution}, probability distribution
\eqref{eq:NBD} is a special case of \emph{Compound Poisson Distribution},
so that \eqref{eq:NBD} describes random composite events (like those
shown in Figure~\ref{fig:CPD}). It was established in \cite{Compound_Distribution}
that these composite events obey Poisson statistics
\begin{equation}
g_{k}(\tau)=\dfrac{(\eta\tau)^{k}}{k\,!}e^{-\eta\tau},\label{eq:Cluster_Statistics}
\end{equation}
where $\eta$ is the average number of composite events per unit volume
\begin{equation}
\eta=\ln\dfrac{1}{p},\label{eq:Feller_eta}
\end{equation}
while probability $f_{k}$, that a composite event consists of $k$
elementary events, is given by the logarithmic distribution 
\begin{equation}
f_{k}=\dfrac{(1-p)^{k}}{k\eta}.\label{eq:Feller_fk}
\end{equation}
In our notations, due to \eqref{eq:p(w)}, equations \eqref{eq:Feller_eta}-\eqref{eq:Feller_fk}
take the form
\begin{equation}
\eta=\ln(1+w),\label{eq:BE_eta}
\end{equation}
\begin{equation}
f_{k}=\dfrac{w^{k}}{k(1+w)^{k}\ln(1+w)}.\label{eq:BE_fk}
\end{equation}

The above results, that were obtained in \cite{Compound_Distribution}
from the general theory of Compound Poisson Distribution with respect
to~\eqref{eq:NBD}, are derived in \emph{Appendix~1} from the BE
statistics without resorting to Compound Poisson Distribution. Such
a new derivation provides an independent confirmation of these results.

In the case of quantum statistics, \eqref{eq:BE_eta} gives the average
number of photon clusters per coherence volume (if no distinction
is made between clusters of different ranks), while \eqref{eq:BE_fk}
is the probability that a photon cluster consists of $k$ photons,
$k=1,\,2,\,3,\,...$~.

\section{COMPOSITION OF DISTRIBUTIONS\label{sec:Composition of Distributions}}

The BE statistics is thus the result of superposition of two random
processes, first, a random number of photon clusters entering given
phase-space volume, second, a random number of photons contained in
each cluster. The first random process is described by the Poisson
statistics~\eqref{eq:Cluster_Statistics} while the second process
is described by the logarithmic distribution~\eqref{eq:BE_fk}. 

Such a superposition of two random processes is termed a \emph{composition
of distributions} and entails certain relation between the generating
functions of the respective statistics. 

Let us denote by $P(z)$ the generating function of the BE statistics~\eqref{eq:BE_in_tau_bins}.
By definition of generating function,
\begin{equation}
P(z)=\sum_{n=0}^{\infty}p_{n}(\tau)z^{n}=\dfrac{1}{(1+w-wz)^{\tau}}.\label{eq:GF_P(z)}
\end{equation}
The generating function $G(z)$ of Poisson statistics \eqref{eq:Cluster_Statistics} is 
\begin{equation}
G(z)=\sum_{k=0}^{\infty}g_{k}(\tau)z^{k}=e^{\eta\tau(z-1)}.\label{eq:GF_G(z)}
\end{equation}
For generating function $F(z)$ of logarithmic distribution \eqref{eq:BE_fk}
we obtain:
\begin{equation}
F(z)=\sum_{k=1}^{\infty}f_{k}z^{k}=\frac{1}{\eta}\sum_{k=1}^{\infty}\frac{(bz)^{k}}{k}=\dfrac{\ln(1-bz)}{\ln(1-b)},\label{eq:GF_F(z)}
\end{equation}
where a notation is introduced 
\begin{equation}
b=\dfrac{w}{1+w}.\label{eq:b(w)}
\end{equation}

The BE statistics is the result of composition of distributions \eqref{eq:Cluster_Statistics}
and \eqref{eq:BE_fk}, so the generating functions of the statistics
involved should satisfy~\cite{Compound_Distribution}
\begin{equation}
P(z)=G\left[F(z)\right].\label{eq:P=00003DG(F)}
\end{equation}

Substituting \eqref{eq:GF_P(z)}, \eqref{eq:GF_G(z)} and \eqref{eq:GF_F(z)}
in \eqref{eq:P=00003DG(F)} we can verify that the latter equation
becomes an identity 
\begin{equation}
\dfrac{1}{(1+w-wz)^{\tau}}\equiv\exp\left[\eta\tau\left(\dfrac{\ln(1-bz)}{\ln(1-b)}-1\right)\right].\label{eq:Identity}
\end{equation}
Indeed, given \eqref{eq:b(w)} and \eqref{eq:BE_eta} the right side
of \eqref{eq:Identity} can be presented as
\[
\exp\left[-\frac{\eta\tau\ln\left(1+w-wz\right)}{\ln\left(1+w\right)}\right]=\exp\left[\ln\left(1+w-wz\right)^{-\tau}\right],
\]
which is identical to the left side of~\eqref{eq:Identity}. 

This result for generating functions is another confirmation of correctness
of equations \eqref{eq:Cluster_Statistics}, \eqref{eq:BE_eta} and
\eqref{eq:BE_fk} in the BE statistics.

\section{PHOTON CLUSTER RADIATION SPECTRA IN BLACKBODY CAVITY\label{sec:Cluster radiation spectra}}

The result of measuring thermal radiation energy in a narrow frequency
range will depend on what is actually measured -- the number of photons
(i.\,e. the total radiation energy), or the number of clusters. Total
radiation energy can be measured, for example, by a bolometer, while
the number of clusters can be measured by a photomultiplier tube since
a single cluster should produce a single click in the photomultiplier.
Therefore, it is necessary to distinguish between the radiation spectrum
measured by the bolometer and the spectrum measured by photomultiplier.
In this section, we will focus on the energy spectrum of thermal radiation
measured by a bolometer that, by assumption, absorbs the total radiation
energy within the selected narrow frequency interval.

\subsection{Mode-average number of photons in the same-rank cluster radiation\label{sub:k_m}}

Thermal radiation in a blackbody cavity is usually presented as a
sum of standing waves, or modes. Each radiation mode of certain frequency,
polarization and direction of propagation corresponds to a single
phase-space cell. The Heisenberg uncertainty principle allows one
to associate volume of $h^{3}$ with a single cell, or mode, in six-dimensional
phase-space. This volume projects on the coherence volume in three
dimensional space. 

Let us find the mode-average number of photons $k_{m}$ belonging
to the $m${\small \nobreakdash-}th rank clusters. Using the average
number of clusters $\eta$ per mode and probability $f_{m}$ that
an arbitrary cluster consists of $m$ photons, we obtain for the average
number of $m${\small \nobreakdash-}th rank clusters in a mode 
\begin{equation}
\eta_{m}=\eta f_{m}.\label{eq:eta_m}
\end{equation}
With \eqref{eq:BE_eta} and \eqref{eq:BE_fk} taken into account,
eq.~\eqref{eq:eta_m} yields: 
\begin{equation}
\eta_{m}=\dfrac{1}{m}\left(\dfrac{w}{1+w}\right)^{m}.\label{eq:eta_m(w)}
\end{equation}
This expression is always less than unity $\eta_{m}<1$ since $m=1,\,2,\,3,\,...$.
Therefore, most often, for any integer $m\geq1$ there are no $m${\small \nobreakdash-}th
rank clusters in a mode and only rarely there is one or several such
clusters.

Now for the mode-average number of photons $k_{m}$ that belong to
the $m${\small \nobreakdash-}th rank clusters we obtain 
\begin{equation}
k_{m}=m\eta_{m}=\left(\dfrac{w}{1+w}\right)^{m},\label{eq:k_m}
\end{equation}
which is also always less than unity. Using the known expression for
the degeneracy parameter $w$ in the BE statistics (see derivation
of~\eqref{eq:Mean_ph_number} in Appendix~1) 
\begin{equation}
w=\dfrac{1}{\exp(\beta\varepsilon)-1},\label{eq:BE_mean_ph_number}
\end{equation}
we obtain for \eqref{eq:k_m} 
\begin{equation}
k_{m}=e^{-m\beta\varepsilon},\label{eq:Mean_ph_in_mth_rank_clusters}
\end{equation}
where $\varepsilon$ is the energy of single photon, $\beta=\tfrac{1}{kT}$,
and $m\varepsilon$ is the energy of $m${\small \nobreakdash-}th
rank cluster.

Multiplying the mode-average number of photons belonging to the $m${\small \nobreakdash-}th
rank clusters \eqref{eq:Mean_ph_in_mth_rank_clusters} by the energy
$\varepsilon$ of a single photon we obtain 
\begin{equation}
\varepsilon_{m}=\varepsilon k_{m}=\varepsilon\exp\left(-m\beta\varepsilon\right),\label{eq:Mean_energy_in_mth_rank_clusters}
\end{equation}
which is the mode-average energy carried by $m${\small \nobreakdash-}th
rank clusters.

\subsection{Spectra of blackbody cluster radiation \label{sub:Cluster_Spectra}}

The mode-average energy of $m${\small \nobreakdash-}th rank cluster
radiation \eqref{eq:Mean_energy_in_mth_rank_clusters} times mode
density yields energy density contained in $m${\small \nobreakdash-}th
rank cluster radiation. Before proceeding with this calculation, we
have to realize what is the mode density in case of photon cluster
radiation. 

We do not discuss here the mechanism of photon clusterization in blackbody
radiation. In HOM-type experiments (for example, in~\cite{HOM,Half_lambda,Quarter_lambda}),
indistinguishable photons stick together when they are scattered into
two different modes by a beamsplitter. In the blackbody cavity, indistinguishable
photons stick together, most probably, due to multiple scattering
on the cavity walls. However, the final result of clusterization process,
irrespective of its mechanism, is clearly derived from the BE statistics.
With that it is important to note that when photons stick together
their energy does not change, and hence their frequency remains unchanged.
Therefore, a photon cluster belongs to the same frequency range $\Delta\nu$
that contained photons before sticking. For this reason, when photons
stick together, their coherence length $\Delta L=\tfrac{c}{\Delta\nu}$
also remains unchanged. Hence, a photon cluster coherence volume is
the same as that of constituent photons. Likewise, the same is the
cluster mode density because it is simply the number of coherence
volumes per unit volume.

Therefore, when calculating the mode density of photon cluster radiation,
we can use the standard expression for mode density in a cavity 
\begin{equation}
\Delta N(\varepsilon)=\dfrac{8\pi\varepsilon^{2}}{c^{3}h^{3}}\Delta\varepsilon,\label{eq:Mode_density}
\end{equation}
which refers to the energy interval $\Delta\varepsilon=h\Delta\nu$
with two polarizations taken into account. 

Now multiplying \eqref{eq:Mean_energy_in_mth_rank_clusters} by \eqref{eq:Mode_density}
we obtain the energy density attributable to the $m${\small \nobreakdash-}th
rank photon cluster radiation at frequency~$\nu$: 
\begin{equation}
u_{m}(\nu)=\varepsilon_{m}\Delta N(\varepsilon)=\dfrac{8\pi h\nu^{3}}{c^{3}}\exp\left(-\dfrac{mh\nu}{kT}\right).\label{eq:En_density_in_mth_rank_clusters}
\end{equation}

This formula, written for a unit frequency interval $\Delta\nu=1$,
solves the problem of finding emission spectra of photon clusters
of various ranks in the blackbody cavity.

Note that a photon cluster may have only two polarizations just like
a single photon, which is taken into account in equation~\eqref{eq:Mode_density}.
This is due to the fact that only indistinguishable photons, which
have the same momentum (in magnitude and direction) and the same polarization,
may stick together \cite{HOM,Kaige Wang}. Therefore, all photons
contained in a cluster must have the same polarization.

Summing up the energy density of photon cluster radiation \eqref{eq:En_density_in_mth_rank_clusters}
as a geometric series over all possible ranks $m=1,\,2,\,\ldots$
yields total radiation energy density 
\begin{alignat}{1}
U(\nu)=\sum_{m=1}^{\infty}u_{m}(\nu) & =\frac{8\pi h\nu^{3}}{c^{3}}\sum_{m=1}^{\infty}\exp\left(-\dfrac{mh\nu}{kT}\right)\nonumber \\
 & =\frac{8\pi h\nu^{3}}{c^{3}}\frac{1}{e^{\nicefrac{h\nu}{kT}}-1}.\label{eq:Planck_s_law(nu)}
\end{alignat}

Equation \eqref{eq:Planck_s_law(nu)} is the standard Planck's formula
for radiation energy density in the blackbody cavity. Hence, considerations
presented above actually constitute a new method to derive the Planck's
radiation law \eqref{eq:Planck_s_law(nu)} based on the concept of
cluster nature of blackbody radiation. Equation \eqref{eq:En_density_in_mth_rank_clusters}
is, therefore, a generalization of the Planck's radiation law for
thermal radiation of $m${\small \nobreakdash-}th rank photon clusters.

\begin{figure}[ht]
\noindent \begin{centering}
\includegraphics[width=8.5cm]{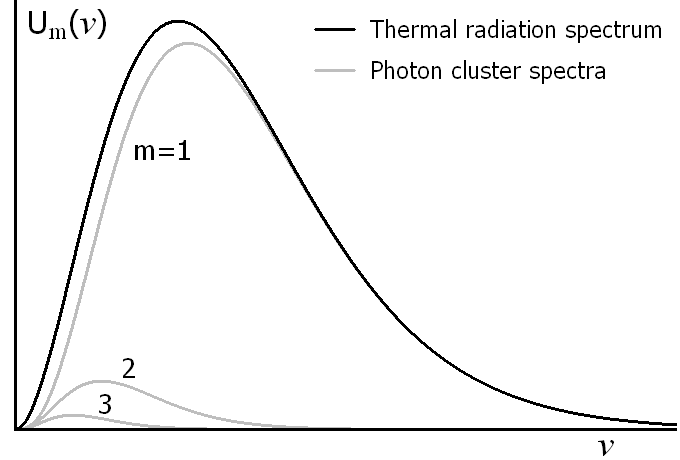}
\par\end{centering}

\caption{\label{fig:nu-spectra}{\small Thermal radiation spectra \eqref{eq:En_density_in_mth_rank_clusters}
of photon clusters: 1\protect\nobreakdash-single photons, 2\protect\nobreakdash-two-photon
clusters, 3\protect\nobreakdash-three-photon clusters. The sum of
all cluster spectra (black curve) yields the Planck's law for blackbody
radiation spectrum \eqref{eq:Planck_s_law(nu)}. The relative position
of curves 1\protect\nobreakdash-2\protect\nobreakdash-3 is temperature-independent.}}
\end{figure}

Spectra \eqref{eq:En_density_in_mth_rank_clusters} for cluster radiation
energy and \eqref{eq:Planck_s_law(nu)} for total radiation energy
are shown in Fig.~\ref{fig:nu-spectra}.

\subsection{The Wien's displacement law for cluster radiation\label{sub:Wien_law}}

Equating to zero the derivative of \eqref{eq:En_density_in_mth_rank_clusters}
with respect to frequency we obtain the condition for maximum of cluster
radiation energy density, which is attained at the frequency 
\begin{equation}
\nu_{m}=\dfrac{3k}{mh}T.\label{eq:Wien_law_for_clusters}
\end{equation}
This equation is the Wien's displacement law prototype for the $m$\nobreakdash-th
rank cluster radiation. With the increase of rank $m$ the maximum
photon cluster radiation energy shifts to a lower frequency. Therefore,
clusters of high ranks are emitted mostly at low frequencies, which
is also obvious from Fig.~\ref{fig:nu-spectra}.

\subsection{The Stefan-Boltzmann law for cluster radiation\label{sub:Stefan-Boltzmann_law}}

Integrating \eqref{eq:En_density_in_mth_rank_clusters} over frequency,
we obtain the total energy density of thermal radiation of $m$\nobreakdash-th
rank clusters
\begin{equation}
U_{m}=\sigma_{m}T^{4},\label{eq:St_B_law_for_mth_rank_cl}
\end{equation}
which is the Stefan-Boltzmann law prototype for photon cluster radiation.
The constant of proportionality in \eqref{eq:St_B_law_for_mth_rank_cl}
depends on the rank $m$ of cluster radiation: 
\begin{equation}
\sigma_{m}=\dfrac{48\pi k^{4}}{m^{4}c^{3}h^{3}}.\label{eq:S_B_const_for_clusters}
\end{equation}

Obviously, the total energy in cluster radiation decreases sharply
with the increase of the cluster rank. Therefore, blackbody radiation
energy is carried mostly by single photons, i.\,e. clusters of rank
$m=1$. Summing \eqref{eq:S_B_const_for_clusters} over all ranks
$m$, we obtain the usual Stefan-Boltzmann constant: 
\begin{equation}
\sigma=\sum_{m=1}^{\infty}\sigma_{m}=\dfrac{8\pi^{5}k^{4}}{15c^{3}h^{3}}\label{eq:St_B_Const}
\end{equation}
where we took into account the known series sum 
\begin{equation}
\sum_{m=1}^{\infty}\dfrac{1}{m^{4}}=\frac{\pi^{4}}{90}.\label{eq:pi^4}
\end{equation}

It follows from \eqref{eq:S_B_const_for_clusters} that the areas
under the curves $m=1,\,2,\,3$ in Figure~\ref{fig:nu-spectra} are
related as $1:\tfrac{1}{16}:\tfrac{1}{81}$.

\subsection{The clusterization degree in blackbody radiation\label{sub:Cl_Degree}}

Let us find the portion $\phi$ of total radiation energy that is
carried by single photons. Using \eqref{eq:S_B_const_for_clusters}
and \eqref{eq:St_B_Const}, we obtain
\begin{equation}
\phi=\frac{\sigma_{1}}{\sigma}=\frac{90}{\pi^{4}}\simeq0.9239,\label{eq:Ph_portion}
\end{equation}
in other words, about 92.4\% of the total energy of blackbody radiation
is carried by single photons. Two-photon clusters, according to \eqref{eq:S_B_const_for_clusters},
account for 16 times less energy, or about 5.77\% of the total energy
of thermal radiation. 

Let us define the degree of clusterization of blackbody radiation
as
\begin{equation}
\Theta=\frac{\sigma-\sigma_{1}}{\sigma}=1-\frac{90}{\pi^{4}}\simeq0.0761,\label{eq:Clusterization_degree}
\end{equation}
which means that all multiphoton clusters (starting with two-photon
clusters) transfer about 7.6\% of thermal radiation energy. 

The clusterization degree \eqref{eq:Clusterization_degree} of thermal
radiation is a universal constant that does not depend on the blackbody
temperature, nor does it depend on physical constants, such as the
Planck's constant or the speed of light. 

In Fig.~\ref{fig:nu-spectra}, the energy of all photon clusters
with ranks $m\geq2$ corresponds to the area enclosed between curve
\emph{$m=1$} and the Planck's spectrum, which is shown by the solid
black curve. 

It is noteworthy that the energy density of cluster radiation of arbitrary
rank, according to \eqref{eq:En_density_in_mth_rank_clusters}, tends
to a constant $u_{m}(\nu)\rightarrow8\pi h\nu^{3}/c^{3}$
at any fixed frequency as the temperature increases to infinity $T\rightarrow\infty$.
This constant is independent of cluster rank. Therefore, at high temperature
$kT>mh\nu$, cluster radiation of any rank less than $m=\frac{kT}{h\nu}$
has, by the order of magnitude, the same energy density. The total
energy density of thermal radiation tends to infinity as $T\rightarrow\infty$
only because the number of ranks of clusters effectively emitted at
the selected frequency is increasing.

\subsection{Portion of coalescent photons in thermal radiation\label{sub:Portion}}

Dividing the energy density of $m$\nobreakdash-photon cluster radiation
\eqref{eq:En_density_in_mth_rank_clusters} by $\varepsilon=h\nu$
yields the density of photons involved in the creation of $m$\nobreakdash-photon
clusters per unit frequency interval:
\begin{equation}
\rho_{m}(\nu)=\frac{u_{m}(\nu)}{h\nu}=\dfrac{8\pi\nu^{2}}{c^{3}}\exp\left(-\dfrac{mh\nu}{kT}\right).\label{eq:Ph_density_in_m-clusters}
\end{equation}
Integrating \eqref{eq:Ph_density_in_m-clusters} over frequency yields
the total number of photons per unit volume that belong to $m$\nobreakdash-photon
clusters:
\begin{equation}
N_{m}=\int_{0}^{\infty}\rho_{m}(\nu)d\nu=\frac{16\pi k^{3}}{m^{3}c^{3}h^{3}}T^{3},\label{eq:Ph_number}
\end{equation}
which is the Stefan-Boltzmann law prototype for the total number of
photons in $m${\small \nobreakdash-}photon clusters. Wherefrom,
by analogy with \eqref{eq:Clusterization_degree}, one may calculate
the portion of photons involved in photon clusters of ranks~$m\geq2$:
\begin{equation}
\dfrac{N-N_{1}}{N}\simeq0.168,\label{eq:N_portion}
\end{equation}
where $N$ is the total number of photons per unit volume in the blackbody
cavity. It can be evaluated using \eqref{eq:Ph_number} with the zeta-function
appearing in the result: 
\begin{equation}
N=\sum_{m=1}^{\infty}N_{m}=N_{1}\zeta(3)\simeq1.2021\, N_{1}.\label{eq:Zeta(3)}
\end{equation}
It follows from \eqref{eq:N_portion} that almost 17\% of all photons
in the Universe are coalescent, i.\,e. belong to photon clusters
of ranks~$m\geq2$.

\section{PHOTON CLUSTER EMISSION SPECTRA VERSUS WAVELENGTH\label{sec:Cl_spectra_vs_lambda}}

The existence of photon clusters in blackbody radiation results in
somewhat unexpected conclusions regarding the spectrum of blackbody
radiation as a function of wavelength -- the spectrum turns out to
be sensitive to photon clusterization.

Indeed, in order to switch from the frequency to the wavelength in
the Planck's formula \eqref{eq:Planck_s_law(nu)}, one typically makes
use of the relationship between the wavelength and frequency of electromagnetic
radiation: 
\begin{equation}
\nu=\frac{c}{\lambda},\qquad\Delta\nu=\frac{c}{\lambda^{2}}\Delta\lambda.\label{eq:Nu_to_lambda}
\end{equation}
Substituting \eqref{eq:Nu_to_lambda} in \eqref{eq:Planck_s_law(nu)}
gives the standard Planck's law for the energy density in blackbody
cavity as a function of wavelength ($\lambda$\nobreakdash-spectrum):
\begin{equation}
U_{p}(\lambda)=\frac{8\pi hc}{\lambda^{5}}\frac{1}{\exp\left(\dfrac{hc}{\lambda kT}\right)-1}.\label{eq:Planks_law(lambda)}
\end{equation}

However, \eqref{eq:Nu_to_lambda} is valid only for single photons,
and invalid for coalescent photons. This is because the wavelength
$\lambda$ of a quantum particle is inversely proportional to its
momentum $p$ in accordance with the de~Broglie formula $\lambda=\tfrac{h}{p}$.
Therefore, the wavelength $\lambda_{m}$ of $m$\nobreakdash-photon
cluster is related to the wavelength $\lambda$ of a constituent photon~as
\begin{equation}
\lambda_{m}=\dfrac{h}{mp}=\dfrac{\lambda}{m},\label{eq:Lambda_m}
\end{equation}
where $p$ is the momentum of single photon, $mp$ is the momentum
of $m$\nobreakdash-photon cluster that consists of $m$ indistinguishable
photons with the same momentum. Hence, if $m$ photons stick together
forming an $m$-photon cluster then, according to~\eqref{eq:Lambda_m},
their wavelength becomes $m$~times smaller. 

This conclusion is of fundamental importance for the results discussed
below. Equation \eqref{eq:Lambda_m} has been repeatedly confirmed
in experiments with clusters of various ranks \cite{Half_lambda,Quarter_lambda},
so it can be considered as reliably established. It follows from \eqref{eq:Lambda_m}
that if photons stick together to make a photon cluster, it will appear
in a mode with different wavelength. That will inevitably give rise
to some modification of the Planck's formula \eqref{eq:Planks_law(lambda)}.
No modification of Planck's radiation law \eqref{eq:Planck_s_law(nu)}
is required because, in contrast to photon wavelength, photon frequency
is clusterization-insensitive. 

It follows from \eqref{eq:Lambda_m} that $\lambda=m\lambda_{m}$.
Combining this result with \eqref{eq:Nu_to_lambda} we obtain the
following rules:
\begin{equation}
\nu=\frac{c}{m\lambda_{m}},\qquad\Delta\nu=\frac{c}{m\lambda_{m}^{2}}\Delta\lambda_{m}.\label{eq:Nu_to_lambda_for_cl}
\end{equation}
Equations \eqref{eq:Nu_to_lambda_for_cl} take into account the fact
that if $m$ indistinguishable photons stick together then their frequency
does not change while their wavelength becomes $m$ times smaller.
The second equation in \eqref{eq:Nu_to_lambda_for_cl} is obtained
by differentiating the first one. 

Equations \eqref{eq:Nu_to_lambda_for_cl} shall be used instead of
\eqref{eq:Nu_to_lambda} if connection between frequency and wavelength
is needed for the $m${\small \nobreakdash-}photon cluster radiation.
In this respect, \eqref{eq:Nu_to_lambda_for_cl} is the generalization
of \eqref{eq:Nu_to_lambda} for cluster radiation. 

Substituting \eqref{eq:Nu_to_lambda_for_cl} in \eqref{eq:En_density_in_mth_rank_clusters}
for a unit wavelength interval $\Delta\lambda_{m}=1$ and renaming
independent variable $\lambda_{m}$ into~$\lambda$, we obtain: 
\begin{equation}
u_{m}(\lambda)=\dfrac{8\pi hc}{m^{4}\lambda^{5}}\exp\left(-\dfrac{hc}{\lambda kT}\right),\label{eq:Cluster_en_density_vs_lambda}
\end{equation}
which is the energy density of thermal radiation of $m${\small \nobreakdash-}photon
clusters versus wavelength. All functions \eqref{eq:Cluster_en_density_vs_lambda}
for various cluster ranks have the same shape with the only difference
being in amplitude, which is inversely proportional to the fourth
power of cluster rank. Therefore, at any wavelength, energy densities
of thermal radiation of single-photon clusters, two-photon clusters,
and three-photon clusters are related as 1:$\tfrac{1}{16}$:$\tfrac{1}{81}$,
respectively.

Summing \eqref{eq:Cluster_en_density_vs_lambda} over all ranks $m$
with \eqref{eq:pi^4} taken into account we obtain a total radiation
energy density as a~function of wavelength: 
\begin{equation}
U(\lambda)=\sum_{m=1}^{\infty}u_{m}(\lambda)=\dfrac{4\pi^{5}hc}{45\lambda^{5}}\exp\left(-\dfrac{hc}{\lambda kT}\right).\label{eq:Modified_Plank_s_law}
\end{equation}
This formula, in contrast to the standard Planck's law \eqref{eq:Planks_law(lambda)},
takes into account the cluster nature of blackbody radiation. The
standard Planck's law \eqref{eq:Planks_law(lambda)} is valid only
in a single-photon approximation to the radiation field and does not
take into account the existence of photon clusters.

\begin{figure}[ht]
\noindent \begin{centering}
\includegraphics[width=8.5cm]{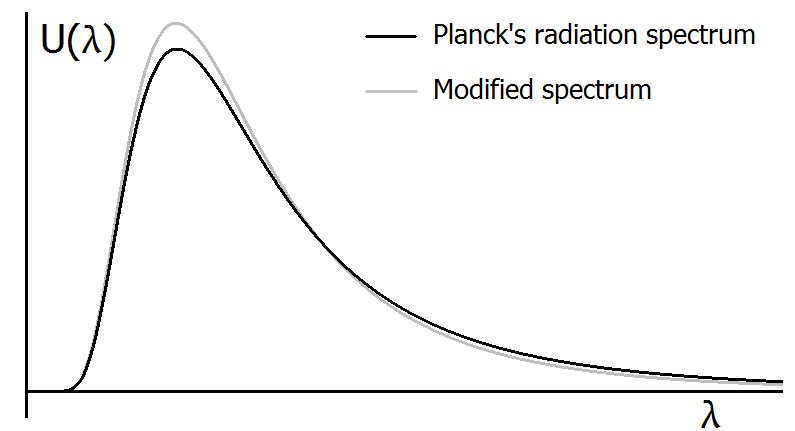}
\par\end{centering}

\caption{\label{fig:Mod_Sp}{\small The Planck's law \eqref{eq:Planks_law(lambda)}
for blackbody radiation spectrum (black) as a function of wavelength
$\lambda$ and modified spectrum \eqref{eq:Modified_Plank_s_law}
that takes into account the existence of photon clusters (gray). Both
curves correspond to blackbody temperature $T=6000K$. }}
\end{figure}

Fig.~\ref{fig:Mod_Sp} shows the modified spectrum \eqref{eq:Modified_Plank_s_law}
in comparison with the standard Planck's law \eqref{eq:Planks_law(lambda)}
thus explaining how the existence of photon clusters affects the $\lambda${\small \nobreakdash-}spectrum
of blackbody radiation.

The modified radiation law \eqref{eq:Modified_Plank_s_law} differs
from the Planck's law \eqref{eq:Planks_law(lambda)} due to the wavelength
reduction in the process of photon clusterization. According to \eqref{eq:Wien_law_for_clusters},
low-frequency photons with a larger wavelength most effectively stick
together. Therefore, radiation energy is pumped through photon clusterization
from the region of larger wavelengths to the region of medium wavelengths.
This process results in the increase of maximum in comparison with
the standard Planck's law (Figure~\ref{fig:Mod_Sp}). The maxima
of two curves are offset by $\sim0.007\lambda$, the ratio of the
two functions at the maximum is approximately equal to $1.075$ while
the areas under the curves coincide exactly.

Thus, if some photons stick together then the standard Planck's law
\eqref{eq:Planks_law(lambda)} for $\lambda${\small \nobreakdash-}spectrum
of thermal radiation must be substituted with equation~\eqref{eq:Modified_Plank_s_law}.

The existence of photon clusters in thermal radiation clarifies the
mechanism of ``photon bunching'' effect that was studied in~\cite{GBD}.
This effect consists in\emph{ }abnormally high probability that photons,
located in an arbitrary volume in blackbody radiation, are found in
a part of that volume. In view of the cluster theory of blackbody
radiation, it is clear that if there is a chance that photons stick
together to form a photon cluster then this cluster will be either
in one or another part of the volume under consideration, thus increasing
the probability that all photons are just in one part of the chosen
volume.

To conclude this Section, the following should be emphasized. If the
BE statistics is valid in the form of the Mandel's formula \eqref{eq:BE_in_tau_bins}
then the blackbody radiation should contain photon clusters and, therefore,
the Planck's law for $\lambda${\small \nobreakdash-}spectrum of
thermal radiation should be modified. If, on the contrary, there are
no photon clusters and the Planck's law is correct then our understanding
of quantum statistics and the Mandel's formula \eqref{eq:BE_in_tau_bins}
should be modified. In other words, the Mandel's formula \eqref{eq:BE_in_tau_bins}
is incompatible with the Planck's law \eqref{eq:Planks_law(lambda)}
because the former predicts the existence of photon clusters while
the latter is valid only in a single-photon approximation to the radiation
field.

\section{CONCLUSIONS\label{sec:Conclusions}}

Fundamentally new results obtained in this paper are based on the
following facts that are firmly established:
\begin{enumerate}
\item The BE statistics in an arbitrary volume is presented by the Mandel's
formula~\eqref{eq:BE_in_tau_bins}.
\item The Mandel's formula is a negative binomial distribution.
\item A negative binomial distribution is a special case of Compound Poisson
Distribution.
\item The wavelength of $N$ coalescent photons is $N$ times smaller than
the wavelength of a single photon.
\end{enumerate}
Facts number 1 and number 3 were established in about the middle of
the 20th century. Fact number 2 is a trivial observation. Fact number
4, which is the consequence of de Broglie's formula, was confirmed
in several independent experiments conducted after 1987 when coalescent
photons were discovered. 

In this paper the following new results are obtained on the basis
of the above facts:
\begin{enumerate}
\item It is proven that the existence of photon clusters (coalescent photons)
is a consequence of quantum statistics.
\item Distribution by rank (which is the number of constituent photons)
is found for photon clusters in blackbody radiation.
\item It is shown that photon clusters in thermal radiation obey the Poisson
statistics.
\item It is proven that the BE statistics is the result of superposition
of two random processes: 1) random number of photon clusters entering
the selected volume, and 2) random number of photons contained in
each cluster. 
\item Spectra of cluster components of blackbody radiation are obtained.
\item It is proven that the Mandel's formula for the BE statistics in an
arbitrary volume is incompatible with the Planck's law for $\lambda${\small \nobreakdash-}spectrum
of thermal radiation. This incompatibility is due to the fact that
the Mandel's formula predicts the existence of photon clusters while
the Planck's formula does not take into account the cluster nature
of radiation.
\item It is shown that the $\lambda${\small \nobreakdash-}spectrum of
 thermal radiation should depend on the level of photon clusterization.
At the same time, photon clusterization does not affect the $\nu${\small \nobreakdash-}spectrum
of thermal radiation, so that the standard Planck's law for the $\nu${\small \nobreakdash-}spectrum
of thermal radiation can be derived from the cluster nature of radiation.
\end{enumerate}
It should be noted that the linearity of Maxwell's equations for electromagnetic
field implies that photons do not interact with each other. Non-interacting
particles must obey Poisson statistics, which is true for photons
in a coherent field. The BE statistics, which is valid for blackbody
radiation, differs from the Poisson statistics. That difference, however,
does not imply that there is any interaction between photons in blackbody
radiation -- there is certainly no interaction. Instead, there is
some probability that photons may be found in coalescent states, or
in the form of photon clusters. That is the reason behind the difference
between the BE statistics and the classical Poisson statistics.

\section*{Appendix 1\label{sec:Appendix-1}}

\begin{center}
\textbf{Photon cluster statistics in thermal radiation and cluster
distribution by the number of constituent photons}
\par\end{center}

\subsection{Photon cluster statistics\label{sub:Cluster_statistics}}

Probability $g_{k}(\tau)$ that $k$ clusters (regardless of their
rank) are in volume~$\tau$ will be found here without resort to
Compound Poisson Distribution.

It follows from \eqref{eq:Rel_Prob_in_BE} that in quantum statistics
relative probability $q_{n}(\tau)$ that $n$ photons are in volume
$\tau$ can be presented as 
\begin{equation}
q_{n}(\tau)=\dfrac{b^{n}}{n\,!}\tau(\tau+1)\ldots(\tau+n-1)\label{eq:Polynom}
\end{equation}
where 
\begin{equation}
b=\frac{w}{1+w}\label{eq:b}
\end{equation}
The rising factorial $\tau^{\bar{n}}$ in \eqref{eq:Polynom} can
be expanded in powers of~$\tau$: 
\begin{equation}
\tau(\tau+1)\ldots(\tau+n-1)=S_{n1}\tau+S_{n2}\tau^{2}+\ldots+S_{nn}\tau^{n}\label{eq:Stirling_numbers}
\end{equation}
where $S_{nk}$ are Stirling numbers of the first kind. 

Introducing notation
\begin{equation}
c_{nk}=\dfrac{b^{n}}{n\,!}S_{nk}\label{eq:c_nk}
\end{equation}
we can present \eqref{eq:Polynom} as a system of equations for {$n=0,\,1,\,2,\,\ldots$} 

\begin{eqnarray}
q_{0}(\tau) & = & 1\nonumber \\
q_{1}(\tau) & = & 0+c_{11}\tau\nonumber \\
q_{2}(\tau) & = & 0+c_{21}\tau+c_{22}\tau^{2}\label{eq:Polynomials}\\
\cdots &  & \cdots\nonumber \\
q_{n}(\tau) & = & 0+c_{n1}\tau+c_{n2}\tau^{2}+\cdots+c_{nn}\tau^{n}\nonumber \\
\cdots &  & \cdots\nonumber 
\end{eqnarray}

Based on the meaning of equations \eqref{eq:1P}{\small \nobreakdash-}\eqref{eq:Two_particles}
discussed in Section~\ref{sec:Basic ideas}, we may conclude that
$c_{nk}\tau^{k}$ in \eqref{eq:Polynomials} is relative
probability that $k$ arbitrary clusters are in volume $\tau$ provided
there are $n$ photons ($n\geq k$) in this volume. A~single photon
is considered to be a one-photon cluster. 

For example, if there are three photons in the selected volume then
$c_{31}\tau$ is the probability that all three photons are stuck
together into one three-photon cluster, $c_{32}\tau^{2}$ is the probability
that there are two clusters in volume~$\tau$ (a two-photon cluster
and a separate photon), and $c_{33}\tau^{3}$ is the probability that
three particles (i.\,e. three separate photons) are in volume~$\tau$.

It follows from the above that the sum of all elements of $k${\small \nobreakdash-}th
column in \eqref{eq:Polynomials} gives the relative probability that
$k$ arbitrary clusters are in volume~$\tau$. We denote this quantity
by~$u_{k}(\tau)$: 
\begin{equation}
u_{k}(\tau)=\dfrac{g_{k}(\tau)}{g_{0}(\tau)}=\tau^{k}\sum_{n=0}^{\infty}c_{nk}.\label{eq:Cluster_statistics_1}
\end{equation}
Obviously, 
\begin{equation}
g_{0}(\tau)=p_{0}(\tau)\label{eq:g_0=00003Dp_0}
\end{equation}
because if there are no photons in volume $\tau$ then there are no
photon clusters in this volume (and vice versa). The summation in
\eqref{eq:Cluster_statistics_1} may start from zero since matrix
$c_{nk}$ is triangular (the first $k$ elements in each column are
zeros).

Given \eqref{eq:c_nk} the sum in \eqref{eq:Cluster_statistics_1}
can be reduced to the known generating function for the Stirling numbers
of the first kind~\cite{Concrete Mathematics}: 
\begin{equation}
\sum_{n=0}^{\infty}c_{nk}=\sum_{n=0}^{\infty}\frac{S_{nk}}{n!}b^{n}=\dfrac{1}{k\,!}\left(\ln\dfrac{1}{1-b}\right)^{k},\label{eq:Sum_Cnk}
\end{equation}
which due to \eqref{eq:b} may be written as
\begin{equation}
\sum_{n=0}^{\infty}c_{nk}=\frac{\ln^{k}(1+w)}{k!}.\label{eq:Sum_Cnk_final}
\end{equation}
Now, from \eqref{eq:Cluster_statistics_1} and \eqref{eq:Sum_Cnk_final}
we obtain for the statistics of photon clusters
\begin{equation}
g_{k}(\tau)=g_{0}(\tau)\frac{\tau^{k}}{k!}\ln^{k}(1+w),\label{eq:Cluster_statistics_2}
\end{equation}
where $g_{k}(\tau)$ is the probability that $k$ arbitrary photon
clusters (regardless of their rank) are in volume~$\tau$.

Note that in view of \eqref{eq:g_0=00003Dp_0} and \eqref{eq:P0_BE}
statistics \eqref{eq:Cluster_statistics_2} can be presented as
\begin{equation}
g_{k}(\tau)=\dfrac{(\eta\tau)^{k}}{k\thinspace!}\exp(-\eta\tau),\label{eq:Poisson}
\end{equation}
which is a Poisson statistics with parameter 
\begin{equation}
\eta=\ln(1+w).\label{eq:Eta}
\end{equation}
So we obtained results \eqref{eq:Cluster_Statistics} and \eqref{eq:BE_eta},
which have long been known in the theory of Compound Poisson Distribution~\cite{Compound_Distribution}.

It follows from \eqref{eq:Poisson} that $\eta$ is the average number
of clusters (of any rank) in a coherence volume. Parameter $\eta$
is always less than the mode-average number of photons $w$ because
some photons are stuck together to form a photon cluster. Hence, for
the average number of photons per cluster we obtain 
\begin{equation}
\frac{w}{\eta}=\frac{w}{\ln(1+w)}.\label{eq:Mean_Ph_per_Cluster}
\end{equation}

This quantity is always greater than unity. In the high-frequency
part of the spectrum, the degeneracy parameter is small $w\ll1$.
In this limit, as it follows from \eqref{eq:Mean_Ph_per_Cluster},
there is about one photon per cluster. This means that in the UV limit,
thermal radiation contains a vanishingly small number of multiphoton
clusters. However, in the low-frequency limit, where $w\gg1$, one
cluster may contain a large number of photons. For example, if $w=50$
then, according to \eqref{eq:Mean_Ph_per_Cluster}, one cluster contains
on the average over 12 photons. Thus, in the low-frequency part of
the spectrum, thermal radiation consists mostly of multiphoton clusters.

This situation is typical only for thermal radiation, as well as for
the states of radiation field close to thermal equilibrium. There
is every reason to believe that if the radiation source is far from
thermodynamic equilibrium then the portion of multiphoton clusters
in its radiation is small, while in the emission of strongly non-equilibrium
light sources, such as lasers, multiphoton clusters are absent and
all the clusters contain just a single photon.

Note that when indistinguishable photons stick together their energy
does not change because there is no interaction between photons. Therefore,
the clusterization process cannot affect the partition function $Z$
of photon gas: 
\begin{equation}
Z=\sum_{n=0}^{\infty}e^{-n\beta\varepsilon}=\dfrac{1}{1-\exp\left(-\beta\varepsilon\right)},\label{eq:Partition_function}
\end{equation}
where $\beta=\nicefrac{1}{kT}$, and $\varepsilon=h\nu$ is the energy
of one-photon excitation of radiation mode at frequency~$\nu$. If
the partition function is not sensitive to photon clusterization then
the results obtained from the partition function are valid regardless
of clusterization process. In particular, for the mean energy of mode
excitation, taking into account \eqref{eq:Partition_function}, we
obtain 
\begin{equation}
\left\langle \varepsilon\right\rangle =-\dfrac{1}{Z}\dfrac{\partial Z}{\partial\beta}=\dfrac{\varepsilon}{\exp\left(\beta\varepsilon\right)-1},\label{eq:Mean_energy}
\end{equation}
from where the average number of photons per mode is 
\begin{equation}
w=\dfrac{\left\langle \varepsilon\right\rangle }{h\nu}=\dfrac{1}{\exp\left(\beta\varepsilon\right)-1}.\label{eq:Mean_ph_number}
\end{equation}

Equations \eqref{eq:Eta} and \eqref{eq:Mean_ph_number} yield
\begin{equation}
\eta=-\ln\left[1-\exp\left(-\beta\varepsilon\right)\right],\label{eq:eta(nu,T)}
\end{equation}
which is the mean number of photon clusters per mode as a function
of the radiation frequency and blackbody temperature. Comparing \eqref{eq:eta(nu,T)}
with \eqref{eq:Partition_function} we may conclude that the mode-average
number of clusters 
\begin{equation}
\eta=\ln Z\label{eq:lnZ}
\end{equation}
is simply the logarithm of the partition function of the photon gas.

\subsection{Distribution of photon clusters by rank (by the quantity of constituent
photons)\label{sub:Cluster_rank_distr}}

Let us find the probability $f_{m}$ that a photon cluster chosen
at random consists of $m$ photons. If $\eta$ is the mode-average
number of clusters then $f_{m}\eta$ is the mode-average quantity
of $m${\small \nobreakdash-}photon clusters. Then $mf_{m}\eta$
is the mode-average number of photons contained in the $m${\small \nobreakdash-}photon
clusters. Therefore, the sum of $mf_{m}\eta$ over all ranks $m$
should give the mode-average number of photons: 
\begin{equation}
\sum_{m=1}^{\infty}mf_{m}\eta=w.\label{eq:W_from_clusters}
\end{equation}
An infinite number of unknown quantities $f_{m}$ can be found from
equation \eqref{eq:W_from_clusters} because the right-hand side of
this equation is the known expandable function.

Indeed, eq. \eqref{eq:b} gives 
\begin{equation}
w=\dfrac{b}{1-b}.\label{eq:w(b)}
\end{equation}
Comparing this expression with \eqref{eq:Mean_ph_number} we obtain
\begin{equation}
b=e^{-\beta\varepsilon}.\label{eq:b(eps)}
\end{equation}
This implies that $0<b<1$ for any frequency and temperature. Therefore,
\eqref{eq:w(b)} can be expanded as
\begin{equation}
w=\dfrac{b}{1-b}=b+b^{2}+\ldots+b^{k}+\ldots\label{eq:w(b+)}
\end{equation}
In equation \eqref{eq:W_from_clusters} we change variables according~to
\begin{equation}
f_{m}=x_{m}b^{m},\label{eq:f_m(x_m)}
\end{equation}
which means that instead of the unknown quantities $f_{m}$ we will
seek for~$x_{m}$.

Substituting \eqref{eq:w(b+)} and \eqref{eq:f_m(x_m)} in~\eqref{eq:W_from_clusters}
yields 
\begin{equation}
\eta\sum_{m=1}^{\infty}mx_{m}b^{m}=b+b^{2}+\ldots+b^{m}+\ldots\,\,.\label{eq:Sum}
\end{equation}
This equation, as well as \eqref{eq:W_from_clusters}, should hold
identically for any~$b$. Therefore, equating the coefficients at
equal powers of $b$ in the right and left sides of \eqref{eq:Sum},
we obtain the unknown coefficients: 
\begin{equation}
x_{m}=\dfrac{1}{m\eta}.\label{eq:Xm}
\end{equation}
Using \eqref{eq:Eta}, \eqref{eq:b}, and \eqref{eq:Xm} we obtain
from~\eqref{eq:f_m(x_m)}: 
\begin{equation}
f_{m}=x_{m}b^{m}=\dfrac{b^{m}}{m\eta}=\frac{w^{m}}{m\left(1+w\right)^{m}\ln(1+w)},\label{eq:f_m(m)}
\end{equation}
This result coincides with \eqref{eq:BE_fk}, i.\,e. the problem
of finding probabilities $f_{m}$ is solved. Equation~\eqref{eq:f_m(m)},
due to~\eqref{eq:Mean_ph_number}, gives distribution of photon clusters
by the number of constituent photons as a function of radiation frequency
and blackbody temperature. 

It is evident from \eqref{eq:f_m(m)} that expression for $f_{0}$
is meaningless because there are no clusters containing zero photons.
Therefore any summation over cluster ranks should start from $m=1$.

Using \eqref{eq:b} and \eqref{eq:Eta} it is easy to verify that
probabilities \eqref{eq:f_m(m)} satisfy the normalization condition
\begin{equation}
\sum_{m=1}^{\infty}f_{m}=\dfrac{1}{\eta}\sum_{m=1}^{\infty}\dfrac{b^{m}}{m}=\dfrac{-\ln(1-b)}{\eta}=1.\label{eq:Normalization_of_f_m}
\end{equation}

The main results of Appendix~1 may be summarized as follows: statistics
of photon clusters \eqref{eq:Poisson} and distribution of clusters
by the number of constituent photons \eqref{eq:f_m(m)} obtained in
Appendix~1 are fully consistent with the results obtained in \cite{Compound_Distribution}
for the Negative Binomial Distribution as a special case of Compound
Poisson Distribution.

\end{document}